\documentclass[letterpaper, 10 pt, conference]{ieeeconf}  

\IEEEoverridecommandlockouts                              

\usepackage[pdftex]{graphicx}
\graphicspath{{img/}}
\DeclareGraphicsExtensions{.pdf,.jpeg,.png}

\usepackage{amsmath} 

\usepackage[caption=false,font=footnotesize]{subfig}
\usepackage{amssymb}

\usepackage{soul, color}
\usepackage{amsfonts, bm}
\usepackage{url}

\title{\LARGE \bf
Grid-Interactive Multi-Zone Building Control \\ Using Reinforcement Learning with Global-Local Policy Search
}

\author{Xiangyu Zhang, Rohit Chintala, Andrey Bernstein, Peter Graf and Xin Jin
\thanks{The authors are with the U.S. National Renewable Energy Laboratory, Golden, CO 80401, USA. {\tt\small andrey.bernstein@nrel.gov}}%

\thanks{This work was authored by the National Renewable Energy Laboratory, operated by Alliance for Sustainable Energy, LLC, for the U.S. Department of Energy (DOE) under Contract No. DE-AC36-08GO28308. Funding provided by the U.S. Department of Energy Office of Energy Efficiency and Renewable Energy Building Technologies Office. The views expressed in the article do not necessarily represent the views of the DOE or the U.S. Government. The U.S. Government retains and the publisher, by accepting the article for publication, acknowledges that the U.S. Government retains a nonexclusive, paid-up, irrevocable, worldwide license to publish or reproduce the published form of this work, or allow others to do so, for U.S. Government purposes.}
\thanks{This research was performed using computational resources sponsored by the Department of Energy's Office of Energy Efficiency and Renewable Energy and located at the National Renewable Energy Laboratory.}
}

\begin{document}

\maketitle
\thispagestyle{empty}
\pagestyle{empty}

\begin{abstract}

In this paper, we develop a grid-interactive multi-zone building controller based on a deep reinforcement learning (RL) approach. The controller is designed to facilitate building operation during normal conditions and demand response events, while ensuring occupants comfort and energy efficiency. We leverage a continuous action space RL formulation, and devise a two-stage global-local RL training framework. In the first stage, a global fast policy search is performed using a gradient-free RL algorithm. In the second stage,  a local fine-tuning is conducted using a policy gradient method. In contrast to the state-of-the-art model predictive control (MPC) approach, the proposed RL controller does not require complex computation during real-time operation and can adapt to non-linear building models. We illustrate the controller performance numerically using a five-zone commercial building.

\end{abstract}

\section{Introduction}

In the U.S., buildings consumed around 39\% of the total national energy usage in 2019 (21\% for residential and 18\% for commercial) \cite{eia2019energy} and over 70\% of the total electricity usage \cite{somasundaram2014reference}.
Specifically, in the commercial sector, heating, ventilation and air-conditioning (HVAC) systems account for 32.7\% of the total electricity consumption \cite{eia2012cbecs}. Due to its strong coupling with the building thermal dynamics, HVAC control is complicated and is of interest in this and many other studies.
Proper HVAC control will not only improve thermal comfort and energy efficiency, but also enable buildings to be grid-interactive, providing valuable demand-side resources in the smart grid paradigm. 

Currently, in the literature, model predictive control (MPC) is the mainstream approach for optimal building control. For instance, it has been used for the building cooling system control and demonstrated cost reduction and efficiency improvement in a real-world application \cite{ma2011model}. Moreover, for larger buildings with relatively complex HVAC systems, a distributed version of MPC is proposed in \cite{hou2017distributed} for better scalability and more manageable computational burden. See review paper \cite{afram2014theory} for more examples. Despite the fact that MPC is popular in academic research, its applications are not well-developed in the real world; this is due to two major drawbacks associated with MPC: 1) \textit{high implementation cost}: the on-demand computation of MPC requires optimization problems to be solved repeatedly within each prediction interval, which poses high requirements on the hardware and software; 2) \textit{high modeling cost}: MPC requires building dynamics models, which are building-specific and typically can only be developed by a professional team. The modeling efforts include accurately modeling the building and properly simplifying (e.g., linearizing) it so that the optimization problem can be solved efficiently. As a result, building owners' interest in implementing optimal building control wanes due to these high costs.

In recent years, with the development of artificial intelligence, reinforcement learning (RL) stands out for its excellent performance in optimal decision-making \cite{mnih2015human}. In building control, RL is also being investigated as an alternative to MPC due to the following merits: 1) The RL control policy can be trained offline and only computationally-inexpensive policy evaluation is conducted during real-time control. 2) Compared with optimization, RL can handle non-linearity and stochasticity more easily. 3) The RL controller can be re-trained periodically to cope with model drift (i.e., change of environment/season or occupants behavior). As a result, many studies now investigate the feasibility of using RL controllers for building control. Deep Q-network (DQN) is applied to implement variable air-flow volume (VAV) control and observe effective cost reduction when compared with a rule-based controller \cite{wei2017deep}. Similarly, the asynchronous advantage actor-critic (A3C) algorithm is utilized in \cite{zhang2019whole} to optimally control the supply water set point in a building's heating system. In contrast to prior works where a discrete action space is used, an RL algorithm called Zap Q-learning is applied in HVAC control considering a continuous control space \cite{raman2020reinforcement}. In addition to the building-centric control objectives, using RL to coordinate multiple HVAC units to achieve a load reduction goal during demand response (DR) events while maintaining thermal comfort is investigated in \cite{zhang2020edge}.

Though existing RL-based methods have proven effective, we have identified the following knowledge gaps/fields of improvement that are not now addressed or are only partially addressed:

\begin{enumerate}
    \item Continuous action space should be considered for a more accurate control and to avoid action discretization, which may impact control performance.
    \item In addition to normal building-centric operation, grid-interactive building control should be encouraged in the future smart grid/smart city paradigm.
    \item Zone control should be investigated to enable coordination among zones, as most commercial buildings have multiple zones.
    \item To facilitate practical implementation, the perfect forecast of exogenous data, or other hard to acquire information, should not be included as controller inputs.
\end{enumerate}

In this paper, we study a practical RL controller for grid-interactive multi-zone buildings, with the above-mentioned knowledge gaps addressed. 
Compared to prior works, the continuous policy search space, more complicated controlled system/objective (i.e., multi zone building and grid service), and the absent of perfect forecast of exogenous information, inevitably pose more difficulties for the RL controller's training. To overcome these challenges, we propose a \emph{two-stage global-local} RL policy search framework. In the first stage, a gradient-free evolution strategy RL (ES-RL) algorithm \cite{salimans2017evolution} is leveraged for fast global policy convergence. In the second stage,  the proximal policy optimization (PPO) algorithm \cite{schulman2017proximal} is utilized for local policy fine-tuning. We illustrate the controller performance numerically using a five-zone commercial building. In particular,  we demonstrate that our two-stage approach outperforms the policies trained using ES-RL or PPO alone, and we compare the control performance with an equivalent baseline MPC in both linear and non-linear implementations.

\section{Problem Formulation} \label{sec-problem-formulation}

Buildings are usually divided into several zones, and each zone has its relatively independent thermal dynamics. In a multi-zone commercial building (thermal zone $i\in\mathcal{N}=\{1,2,...,N\}$), which utilizes a centralized HVAC system for cooling, the HVAC chiller discharge air temperature ($T^{da}$) and the cooling air flow rate to each zone ($\dot m^i, i \in \mathcal{N}$), can be controlled to provide a comfortable indoor environment for occupants. In this paper, we propose an RL controller that can achieve multi-objective control over the control horizon ($t\in\mathcal{T}$): 1) minimizing the HVAC system's energy consumption; 2) maintaining the indoor comfort; and 3) limiting the maximum power consumption below a certain time-variable threshold $\overline P_t$. These objectives are sometimes mutually conflicting, and thus designing a good controller is challenging.

\subsection{HVAC Optimal Control Formulation}

Mathematically, the multi-objective optimal HVAC control can be formulated as follows.

First, the thermal discomfort cost measures the temperature deviation from the predefined comfort band $[\underline{T}_t^i, \overline{T}_t^i]$, which is assumed to increase first linearly and then quadratically:
\begin{equation} \label{eq-thermal-discomfort}
\mathcal{D}(T_t^i) := \left\{\begin{array}{ll}{\max(T_t^i - \overline{T}_t^i, (T_t^i - \overline{T}_t^i)^2)}, & {T_t^i > \overline{T}_t^i} \\
{\max(\underline{T}_t^i - T_t^i, (\underline{T}_t^i - T_t^i)^2)}, & {T_t^i < \underline{T}_t^i} \\
{0}, & {\text {otherwise}}\end{array}\right.
\end{equation}
where $T_t^i$ is the indoor temperature of zone $i$ at step $t$.

Second, the relationship between the HVAC power consumption ($P_t$) and control variables (i.e., $\mathbf{\dot m}_t=[\dot m_t^1, \dot m_t^2, ..., \dot m_t^N]$ and $T_t^{da}$) is given by:
\begin{equation} \label{eq-ac-power}
    P(\mathbf{\dot m}_t, T_t^{da}) = a(T_t^{out} - T_t^{da})\sum_{i \in \mathcal{N}} \dot m_t^i + b\left(\sum_{i \in \mathcal{N}} \dot m_t^i\right)^3 + c
\end{equation}
in which the first term describes the cooling chiller power ($T_t^{out}$ represents outdoor temperature) and the rest depicts the VAV fan power. For each control interval ($\Delta t$), the energy consumption is given by:
\begin{equation} \label{eq-ac-energy}
    \mathcal{E}_t = P(\mathbf{\dot m}_t, T_t^{da})\Delta t.
\end{equation}

Third, to account for DR events, a power limit violation penalty term is given by:
\begin{equation} \label{eq-dr-violation-cost}
\mathcal{V}_t := \left\{\begin{array}{ll}{(P(\mathbf{\dot m}_t, T_t^{da}) - \overline{P}_t)^2}, & {P(\mathbf{\dot m}_t, T_t^{da}) \geq \overline{P}_t} \\
{0}, & {\text {otherwise}}\end{array}\right. .
\end{equation}
During a DR event ($t \in \mathcal{T}_{DR}$), $\overline{P}_t$ is reduced dramatically and the controller needs to reduce $P_t$ accordingly in order to participate in the DR event.

Combining all three objectives together, the optimal control problem is formulated as:

\begin{equation} \label{eq-opt-formulation}
\begin{aligned}
& \underset{\mathbf{\dot m}_t, T_t^{da}}{\operatorname{minimize}} 
& & \sum_{t\in\mathcal{T}} \mathbf{w}_t [\kappa_1 \sum_{i \in \mathcal{N}}\mathcal{D}(T_t^i), \kappa_2 \mathcal{E}_t, \kappa_3 \mathcal{V}_t]^\top \\
& \text{subject to}
& & \underline{T}^{da} \leq T_t^{da} \leq \overline{T}^{da},  \left(\forall t \right)\\
&&& \underline{\dot m}^{i} \leq \dot m_t^i \leq \overline{{\dot m}}^{i},  \left(\forall t , \forall i \right) \\
&&& \mathbf{T}_{t+1} = \mathcal{F} \left( \mathbf{T}_t, \mathbf{\dot m}_t, T_t^{da} , \bm{\Psi}_t\right) \left(\forall t \right).
\end{aligned}
\end{equation}
In (\ref{eq-opt-formulation}), $\kappa_i$ are factors that monetize the three control objectives; $\mathbf{w}_t = [w_t^1, w_t^2, w_t^3]$ is a weighting vector such that $\sum_{i=1}^3 w_t^i=1$; and $\mathcal{F}$ is the building thermal dynamics model represented by an equality constraint, in which $\mathbf{T}_t\in\mathbb{R}^N$ is the temperature of all zones at step $t$ and $\bm{\Psi}_t$ represents all non-controllable exogenous inputs. The details of the building model are discussed next.

\subsection{Building Thermal Dynamics Model} \label{subsec-building-model}

In contrast to optimization-based approaches, RL does not place constraints on the format of $\mathcal{F}$, which can be the actual building, a high-fidelity building simulator (i.e., EnergyPlus \cite{crawley2001energyplus}) or a reduced-order model (ROM). Considering the low learning efficiency and potentially dangerous exploring actions when learning on the actual building, and the high modeling cost and poor scalability of using EnergyPlus, a ROM, as a data-driven model that represents the middle ground of model accuracy, modeling cost, and scalability, is leveraged in this paper.

A linear parametric modeling approach \cite{chintala2015automated} is used to develop a temperature-predicting model for Zone $i$ using an Auto-Regressive models with eXogenous variables (ARX), which is given by:

\begin{equation} \label{eq-arx-model}
T_t^i = \sum_{j=1}^{n_a} a^{ij} T_{t-j}^i + \sum_{j=1}^{n_b} {\mathbf{b}^{ij}}^\top\mathbf{u}_{t-j}^i
\end{equation}
where $\mathbf{u}_{t}^i$ is the input vector, and $a^{ij}, j\in \{1,..., n_a\}$ and $ \mathbf{b}^{ij}, j \in \{1,..., n_b\}$ are the parameters to be determined. Features to be considered to formulate $\mathbf{u}_{t}^i$ are shown in Table \ref{Table: black-box time series data}, where $Q_t^{solar,i}$ and $Q_t^{int,i}$ are given as exogenous inputs and $Q_t^{HVAC,i}$ is given by:

\begin{equation} \label{eq-q-definition}
    Q_t^{HVAC, i} = C_p \dot{m}_t^i (-T_t^{da}+T_t^i),
\end{equation}
in which $C_p = 1.0~kWh/(kg\cdot K)$ representing the specific heat capacity of air. These features are down-selected via feature selection and then the final components of $\mathbf{u}_{t}^i$ is determined. Note by considering temperatures from other zones, the interactions between different zones are taken into account. Different hyper-parameters are also tested, and in this paper, we use $n_a=1$ and $n_b=1$, which can already predict temperature accurately. In addition, though (\ref{eq-arx-model}) is linear, the $ \dot{m}_t^i \cdot T_t^{da}$ term in (\ref{eq-q-definition}) makes the model non-linear with respect to the control variables. Finally, all zone-specific temperature prediction models (\ref{eq-arx-model}) are combined to form the multi-zone building model $\mathcal{F}$ in (\ref{eq-opt-formulation}). Model parameters $a^{ij}$ and $ \mathbf{b}^{ij}$ are derived from system identification (i.e., learned from building historical data).

\begin{table}[]
\caption{List of inputs considered to predict temperature of zone $i$  }
\label{Table: black-box time series data}
\centering
\scalebox{1.0}{
\begin{tabular}{c|c}
\hline
Variable  & Symbol\\
 \hline
Outdoor air temperature &  $T_t^{out}$\\
Delivered cooling &  $Q_t^{HVAC,i}$\\
Solar heat gain  &  $Q_t^{solar,i}$\\
Internal heat gain (reflects occupancy level)  &  $Q_t^{int, i}$\\
Other zone's temperature &  $T_{t-1}^{j}, j \in \mathcal{N},j \neq i$\\
 \hline
\end{tabular}
}
\end{table}

\section{Controller Design}

In this section, we present the design of the proposed RL controller and an MPC baseline. Basics of RL are not discussed here, and interested readers should refer to \cite{sutton2018reinforcement}.

\subsection{Reinforcement Learning Controller (RLC)}  \label{subsec-rlc}

To be solved by RL, the optimal control problem (\ref{eq-opt-formulation}) is first transformed into a Markov decision process (MDP), which is typically represented by a quintuple $(\mathcal{S}, \mathcal{A}, \mathcal{R}, \mathcal{P}, \gamma)$. Except the state transition probability $\mathcal{P}$, which is implicitly determined by the controlled system and the environment, other elements $(\mathcal{S}, \mathcal{A}, \mathcal{R}, \gamma)$ are defined as follow.

First, the RL state and its complete set ($\mathbf{s}_t \in \mathcal{S}$) represent the information that the RL agent requires to make a decision. The state typically contains information regarding the current system status and some other information related to its future evolving trajectory. In this study, $\mathbf{s}_t=[\mathbf{T}_t, \mathbf{T}_t^{out}, \mathbf{E}_t, \overline{\mathbf{P}}_t, t, \mathbf{w}_t]$ is used:

1) $\mathbf{T}_t \in \mathbb{R}^N$ are the zone temperatures at $t$.

2) $\mathbf{T}_t^{out} = [T_{t-K+1}^{out}, ..., T_{t}^{out}] \in \mathbb{R}^K$ is the $T_t^{out}$ trajectory of the previous $K$ steps. In contrast to some prior works, where a forecast of $T_t^{out}$ is used, we use past data and let the RL agent learn to predict future $T_t^{out}$ implicitly, which avoids developing a forecasting module that adds extra complexity.

3) $\mathbf{E}_t = [f_t, sin_t, cos_t]$ approximates the solar radiation and occupancy level to avoid deploying a sophisticated sensor network within the building. The binary $f_t \in \{0, 1\}$ indicates if the day is a weekday and $sin_t, cos_t$ are the sine and cosine representations for time of the day (e.g., 12:00 PM is represented by $[sin(2\pi \cdot 12/24), cos(2\pi \cdot 12/24)]$). This coarse approximation, inevitably, introduces error; but we will demonstrate it is adequate in a building with a regular schedule in Section \ref{sec-case-study}. 

4) $\overline{\mathbf{P}}_t = [\overline{P}_t, ..., \overline{P}_{t+K}] \in \mathbb{R}^K$ is the power limit for the next $K$ steps. This means DR events, if called, are notified $K$ steps before their start, which is realistic (e.g., the capacity bidding program from San Diego Gas \& Electric ``Day-of'' option will notify customers hours before the event starts \cite{sdge2019cbp}).

5) Finally, the current step $t$ and multi-objective weights $\mathbf{w}_t$ are included in $\mathbf{s}_t$ as well.

Second, the definition of action and its complete set are straight-forward: $\mathbf{a}_t = [\dot m_t^1, \dot m_t^2, ... , \dot m_t^N, T_t^{da}] \in \mathcal{A} \subseteq \mathbb{R}^{N+1}$, corresponding to control variables in (\ref{eq-opt-formulation}). Similarly, the immediate reward is defined according to (\ref{eq-opt-formulation}) as well: $r_t = -\mathbf{w}_t [\kappa_1 \sum_{i \in \mathcal{N}}\mathcal{D}(T_t^i), \kappa_2 \mathcal{E}_t, \kappa_3 \mathcal{V}_t]^\top \in \mathcal{R}$ to be maximized. Finally, a discount factor $\gamma = 0.99$ is used.

To sum up, based on the MDP formulation above, an RL agent will learn from experience a control strategy (formally introduced in Section \ref{subsec-rl-preliminaries} as the \emph{policy}) that maximizes the expected episodic discounted reward: $\mathbb{E}(\sum_{t\in\mathcal{T}} \gamma^{t} r_t)$. Note that although this MDP problem is different from the optimal control problem \eqref{eq-opt-formulation}, it approximates it well as will be shown in Section \ref{sec-case-study}. 

\subsection{Model Predictive Controllers (MPC)}  \label{subsec-mpc}

Two versions of MPC, MPC-ROM and MPC-LIN, are developed to solve (\ref{eq-opt-formulation}) based on the ROM developed in Section \ref{subsec-building-model}. The two versions serve as a reference against which the performance of the RLC is compared to. Specifically, MPC-ROM uses the exact ARX model (\ref{eq-arx-model}), meaning the model used to compute the optimal control actions is the same as the simulation model. In contrast, MPC-LIN is based on a linear approximation of (\ref{eq-arx-model}), which is obtained by performing a Taylor series expansion at every time-step. Since MPC-ROM requires solving a non-convex optimization, the computational requirements make it less feasible for real-world applications as the number of control variables increases. Therefore, in this paper, MPC-ROM serves as a best possible reference for the control performance; in contrast, MPC-LIN shows the performance of a more attainable implementation.

It is also worth noting that MPC requires exogenous inputs for the planning horizon. In this paper, perfect forecasts (PF) of $T_t^{out}$, $Q_t^{solar, i}$ and $Q_t^{int, i}$ are provided to MPC-ROM and MPC-LIN, albeit unrealistic. In contrast, RLC only uses historical data of $T_t^{out}$ and coarse approximation (CA) of $Q_t^{solar, i}$ and $Q_t^{int, i}$ for decision making; recall the definition of RL state $\mathbf{s}_t$ in Section \ref{subsec-rlc}. 
Table \ref{table-rlc-mpc-comparison} summarizes the comparison between the RLC and MPC settings.

\begin{table}[]
\caption{Comparison between RLC and MPC}
\centering
\begin{tabular}{c|ccc}
\hline
                                & \textbf{RL} & \textbf{MPC-LIN} & \textbf{MPC-ROM} \\ \hline
\textbf{Exogenous Data}         & CA          & PF               & PF               \\ \hline
\textbf{System Model Knowledge} & None        & Approximate      & Exact            \\ \hline
\end{tabular}
\label{table-rlc-mpc-comparison}
\end{table}

\section{Global-Local Policy Search Scheme}

In this section, a global-local policy search scheme is developed, which marries merits from two different types of RL algorithm.

\subsection{RL Algorithm Preliminaries} \label{subsec-rl-preliminaries}

RL algorithms are generally categorized as value-based or policy-based. Policy gradient methods, as policy-based approaches, directly learn a parameterized control policy ($\pi_{\bm{\theta}}(\mathbf{a}|\mathbf{s})$, in which $\bm{\theta}$ is a parameter vector) that maximizes a performance measure $J(\bm{\theta})$ \cite{sutton2018reinforcement}, which is typically represented by $\mathbb{E}_{\pi_{\bm{\theta}}}(\sum_{t\in\mathcal{T}} \gamma^{t} r_t)$. At each learning iteration, the RL algorithm will generate a stochastic estimation of the gradient (i.e., $\widehat{\nabla} J(\bm{\theta})$)  from experience and then update $\bm{\theta}$ using gradient ascent as in:
\begin{equation} \label{eq-policy-gradient-ascent}
    \bm{\theta}_{t+1} = \bm{\theta}_t + \alpha \widehat{\nabla}_{\bm{\theta}} J(\bm{\theta})
\end{equation}


Two state-of-the-art policy-based model-free RL algorithms, which have proved to have outstanding performance in many benchmark problems, are used in this study. Evolution strategies RL (ES-RL) \cite{salimans2017evolution}, which uses a zero-order gradient approximation to estimate $\widehat{\nabla} J(\bm{\theta})$, can achieve fast \emph{global} policy learning due to its gradient-free nature and the ability to escape local optima. Proximal Policy Optimization (PPO) \cite{schulman2017proximal} considers KL divergence during learning to avoid destructive large policy updates and thus provides reliable \emph{local} learning.

\subsection{Two-stage Global-Local Policy Searching} \label{sec-two-stage-training}

\begin{table*}[]
\caption{Strengths and Weakness of ES-RL and PPO}
\centering
\begin{tabular}{c|l|l}
\hline
     & \multicolumn{1}{c|}{\textbf{ES-RL}}                                                                                                                    & \multicolumn{1}{c}{\textbf{PPO}}                                                                                                                                                    \\ \hline
\textbf{Pros} & \begin{tabular}[c]{@{}l@{}}- Scalable\\ - Gradient-free (no back-propagation, fast)\\ - More likely to escape from local optimum\end{tabular} & \begin{tabular}[c]{@{}l@{}}- Policy update with KL divergence considered (stable)\\ - Gradient-based (better local search ability)\end{tabular}                               \\ \hline
\textbf{Cons} & - Likely to converge only to the \emph{vicinity} of global optimum \cite{duchi2015}                                                                                       & \begin{tabular}[c]{@{}l@{}}- Slower learning (back-propagation, conservative update)\\ - Not scalable ($\mathcal{O}(N^2)$ communication complexity of full gradient info)\\ - Gradient-based methods are prone to be trapped in local optimum \end{tabular} \\ \hline
\end{tabular}
\label{table-comparison-esrl-ppo}
\end{table*}

Despite the good performance of ES-RL and PPO, using either of them individually might suffer from some drawbacks.
Specifically, training an optimal policy using PPO can be slow due to the back-propagation gradient computation, conservative policy update and poor scalability. On the other hand, due to the adoption of the zero-order gradient approximation, ES-RL is likely to converge to the vicinity of the global optimum \cite{duchi2015}. In summary, the strength and weakness of these two algorithms are summarized in Table \ref{table-comparison-esrl-ppo}. Considering that ES-RL and PPO have complementary advantages, we propose a two-stage global-local policy searching scheme to marry the merits from both algorithms. In the first stage, taking advantage of its gradient-free feature and excellent scalability, ES-RL is utilized for fast policy convergence to the vicinity of the global optimum. 
In the second stage, $\pi_{\bm{\theta}}(\mathbf{a}|\mathbf{s})$ is transferred to a PPO learner, which locally fine-tunes it using computed gradient information and pushes the policy closer to the global optimum. 
By utilizing this two-stage framework, the goal is to train an optimal RL controller in a shorter amount of time and with less computational resources when compared with using these algorithms individually. Next, knowledge transfer between two stages is discussed.

\subsection{PPO Learner Warm-Starting}  \label{subsec-ppo-warm-start}

\begin{figure}[]
\centering
\includegraphics[width=0.99\linewidth]{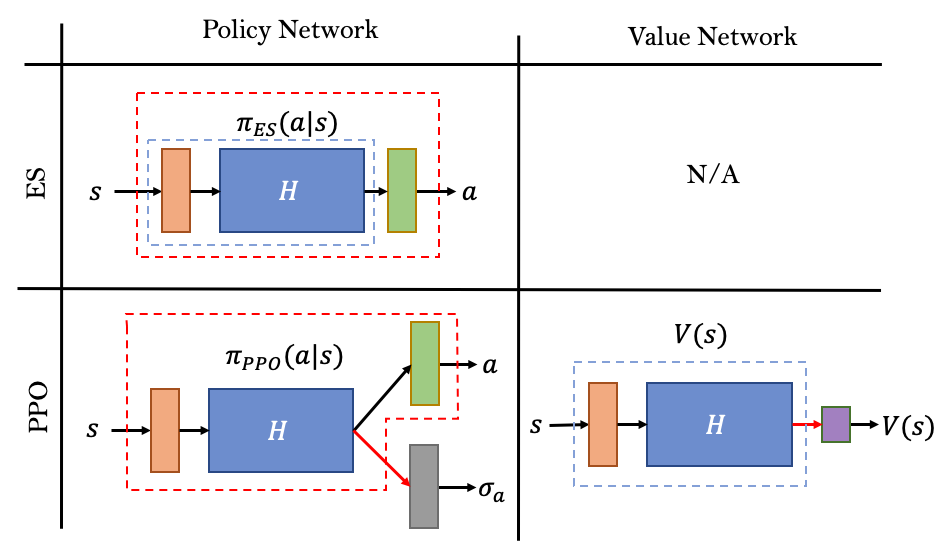}
\caption{Weights of the ES policy network are directly copied to the corresponding portion of the PPO policy network (i.e., red dashed box) to warm-start the PPO learner. Similar weight copying is conducted for the PPO value network (i.e., blue dashed box).}
\label{fig-ppo-warm-start}
\end{figure}

Since ES-RL and PPO both employ neural networks to represent the parameterized policies $\pi_{\bm{\theta}}(\mathbf{a}|\mathbf{s})$, as long as networks share the same/similar structure, the parameters from a trained ES-RL policy can be directly copied to a PPO network ($\bm{\theta}_0^{PPO} = \bm{\theta}^{ES, *}$), which enables the PPO learner to start with nearly as good performance as the well-trained ES learner (though not exactly, as we will explain later). After the weight copying warm start, the PPO learner will further improve $\pi_{\bm{\theta}}(\mathbf{a}|\mathbf{s})$ using a gradient-based approach. Fig. \ref{fig-ppo-warm-start} illustrates the warm-start procedure. It is worth noting that one additional output of the PPO policy network outputs is the standard deviation for each action (denoted as $\mathbf{\sigma_a}$). The PPO learner then samples the actual action at each step by a Gaussian distribution $\mathbf{a}_t = \mathcal{N}(\mathbf{a}, \mathbf{\Sigma})$, where $\mathbf{\Sigma}$ is a diagonal covariance matrix with vector $\mathbf{\sigma_a}$ as diagonal elements. Because neural network weights relate to this additional outputs do not exist in the ES network (i.e., the red arrow in the PPO policy network in Fig. \ref{fig-ppo-warm-start}), we manually initialize the corresponding weights and biases properly to encourage adequate exploration instead of being satisfied with the existing ES policy. Additionally, PPO, as an actor-critic algorithm, uses a value network. We initialize this network using the ES network except for the last layer, due to the different outputs dimension ($V(s)$ is a scalar while $\mathbf{a}$ is a vector). Weights in the last layer are randomly initialized (i.e., red arrow in PPO value network in Fig. \ref{fig-ppo-warm-start}).

\section{Case Study}  \label{sec-case-study}

\subsection{Experiment Setup}

\begin{figure}[]
\centering
\includegraphics[width=0.6\linewidth]{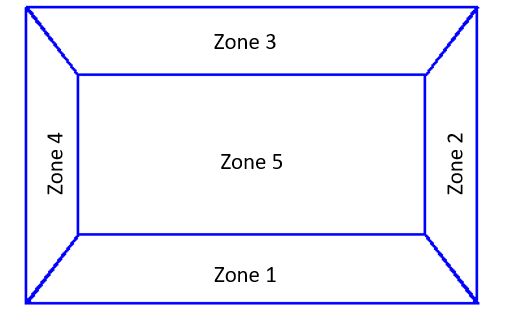}
\caption{A five-zone building widely used in building control studies.}
\label{fig-five-zone-building}
\end{figure}

A standard EnergyPlus five-zone small office model is used in this paper, see Fig. \ref{fig-five-zone-building}. Parameters of the ROM are obtained by system identification using simulated building operation data. The HVAC power consumption model (\ref{eq-ac-power}) has parameters of $a= 1.0$, $b=0.0076$ and $c=4.8865$. Control variables ranges are: $T_t^{da} \in [10.0, 16.0]$ ($^\circ C$), and $\dot m_t^i \in [0.22, 2.2]$ if $i \neq 5$, otherwise, $\dot m_t^5 \in [0.32, 3.2]$ ($kg/s$). We use monetizing factor $\kappa_i = 1.0$ for all three objectives. Exogenous data from EnergyPlus (Austin, TX), including $T_t^{out}$, $Q_t^{solar, i}$ and $Q_t^{int, i}$ ($i\in \mathcal{N}$), from July and the first ten days of August are used for training and controller testing respectively. We formulate the building control problem as a episodic MDP with control horizon length of 24 hours and 5-minute control interval ($\mathcal{T} = \{1, 2, ..., 288\}$). The dimension of $\mathbf{T}_t^{out}$ and $\overline{\mathbf{P}}_t$ is 48 (i.e., $K=48$), which means 1) $T_t^{out}$ for the past 4 hours are used in the state; and 2) a DR event, if called, will be notified 4 hours ahead.

During training, for each episode, we assume the chance a DR event occurs is 50\% (50\% is higher than reality now, but provided with affordable smart controllers, it is assumed that more frequent DR events will be acceptable in the future). DR events happen anytime between 11:00 and 18:00 with a duration of $d$ minutes and a power limit $\overline{P^{DR}}$ kW. The relationship between $d$ and $\overline{P^{DR}}$ is given by:

\begin{equation} \label{eq-duration-and-limit}
    d = 120 \times (\chi + 1), \quad \overline{P^{DR}} = 30 + 20\chi,
\end{equation}
where $\chi$ follows uniform distribution $U(0, 1)$. In real-life applications, this relationship can be changed appropriately. Given the DR starting time and duration, $\mathcal{T}_{DR} \subseteq \mathcal{T}$ is determined; for episodes without DR events, $\mathcal{T}_{DR}=\emptyset$. In an episode, for $t\notin\mathcal{T}_{DR}$ (normal operation), we use $\mathbf{w}_t = [0.7, 0.2, 0.1]$ and $\overline{P}_t=80$kW; and when $t\in\mathcal{T}_{DR}$, we use $\mathbf{w}_t = [0.5, 0.0, 0.5]$ and $\overline{P}_t=\overline{P^{DR}}$.

The policy network $\pi_{\bm{\theta}}(a|s)$ employs the following structure: [108, 256, 128, 128, 64, 64, 32, 16, 6 (ES-RL) / 12 (PPO)] (The first and last items are input and output layers, and the rest show the number of neurons in each hidden layer). Activation functions used in the input and hidden layers are `tanh' and no activation function is used for the output layer. The RL policy training in this study are conducted on a high-performance computing (HPC) system at the U.S. Department of Energy's National Renewable Energy Laboratory (NREL). Choosing an HPC system as the computing hardware is to help evaluate the training cost for future real-world applications, when controllers are trained on commercial cloud computing platforms.

\subsection{Two-stage Policy Searching}

\begin{figure}[]
\centering
\includegraphics[width=0.99\linewidth]{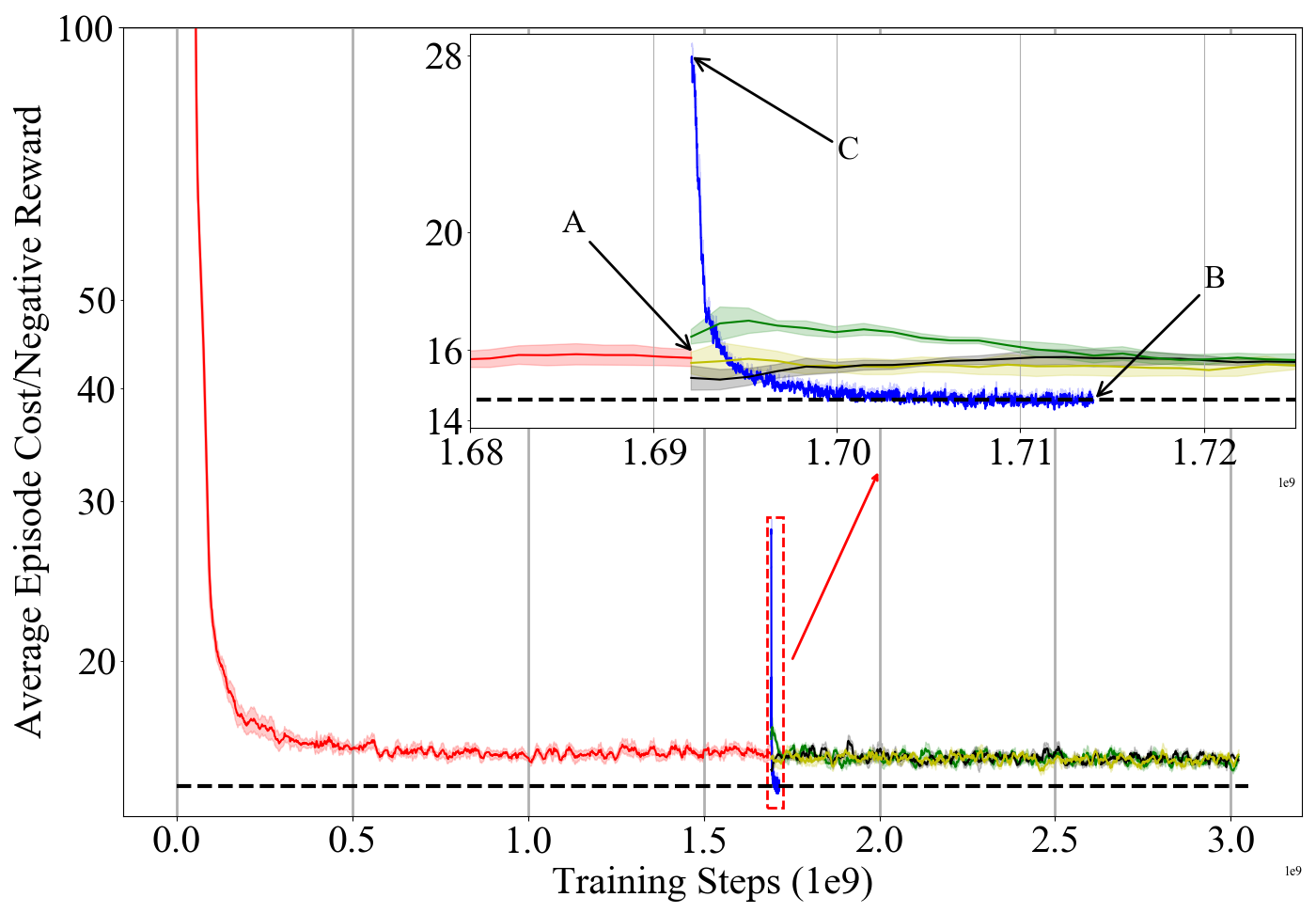}
\caption{Learning curves showing the relationship between average episodic cost and training steps: first stage ES curve (red), second stage PPO fine-tuning curve (blue) and three ES fine-tuning curves with different learning step sizes (green: $5 \times 10^{-6}$, yellow: $10^{-5}$ and black: $10^{-6}$). All curves are averaged from five individual runs and shaded areas show the standard deviation. Red dashed box highlights the PPO fine-tuning, and is zoomed in and shown at the upper right corner. Black horizontal dashed line shows the converged value after PPO fine-tuning.}
\label{fig-two-stage-learning-curves}
\end{figure}

In the first stage, twenty computing nodes with 683 parallel workers (this number is determined by the available processors on computing nodes) are leveraged for the ES-RL global policy search. In Fig. \ref{fig-two-stage-learning-curves}, the red curve shows  effective learning since it reveals a cost-reducing general trend as the training progresses. The minimized cost converged within one-hour of run time (in fact, the convergence occurred within half an hour), implying a converged control policy parameterized by $\bm{\theta}^{ES, *}$. With the trained ES policy network in hand, in the second stage the PPO learner is warm-started as described in Section \ref{subsec-ppo-warm-start} for policy fine-tuning, and the training is conducted by 35 parallel workers on one computing node. The learning curve is shown in blue in Fig. \ref{fig-two-stage-learning-curves}. It is worth noting that though  $\bm{\theta}^{ES, *}$ is copied exactly, the PPO learner starts the average episodic cost from point `C' instead of point `A' in Fig. \ref{fig-two-stage-learning-curves}, showing a slightly deteriorated performance at the beginning. Two reasons for this are: 1) the $\sigma_a$ part of the PPO policy network is initialized to encourage adequate exploration instead of being satisfied with the existing ES policy; thus the PPO learner might explore bad actions as well, which leads to higher cost. 2) the last layer of the PPO value network is randomly initialized, which provides incorrect state evaluation during the first few training iterations. This also leads the PPO learner to have poorer behavior. Despite a start with higher cost, the PPO learner eventually learns a policy with an average episodic cost lower than that achieved by the ES policy, demonstrating an effective policy fine-tuning. When searching the policy globally and locally, learning rates used are $\alpha_1 = 10^{-2}$ and $\alpha_2 = 5 \times 10^{-6}$ respectively.

In this experiment, the two-stage policy search consumes 24 node-hours computational resources (20 nodes for one hour for ES-RL and one node for four hours for PPO) and 5 hours wall time for training the RL controller. Note the first stage convergence happens within the first half hour, it is possible to further reduce the node-hour resources needed.

\subsection{Necessity of the Two-Stage Framework}

In this section, considering reasons presented in Section \ref{sec-two-stage-training}, we provide empirical evidence that the policy trained by the two-stage learning framework cannot be achieved by using either algorithm alone.

\begin{figure}[]
\centering
\includegraphics[width=0.99\linewidth]{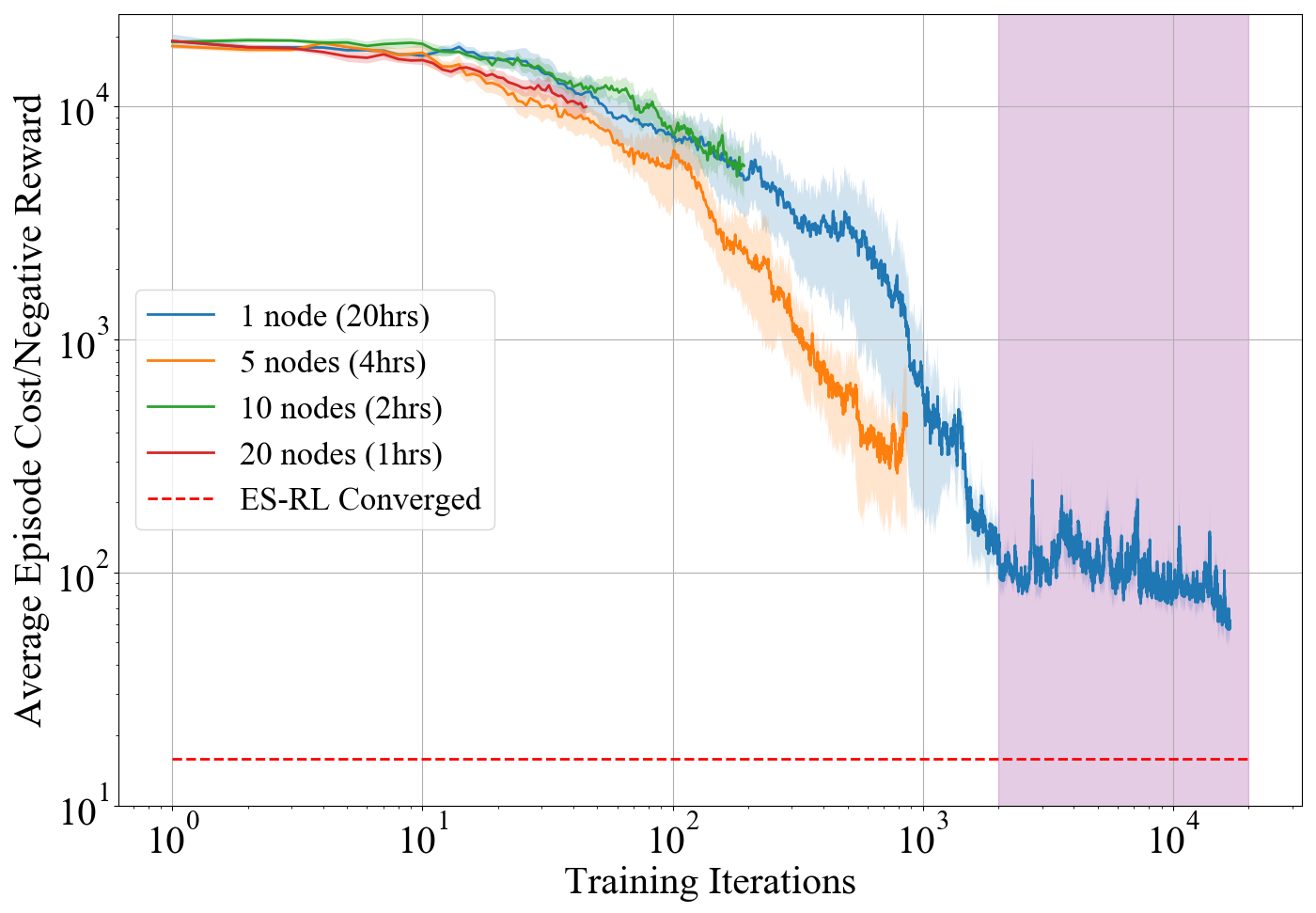}
\caption{Learning curves of learning a control policy using PPO without the first stage ES-RL training. Four learning curves showing the PPO learning process using 1 HPC computing node, 5 nodes, 10 nodes and 20 nodes. All experiments are limited with 20 node-hour of computation resource and the learning rate is equal to the one used in ES-RL (i.e., $\alpha_1$). Again, all curves are averages from five individual runs.}
\label{fig-ppo-from-scratch}
\end{figure}

First, gradient-free ES-RL is not effective for fine-tuning. We tested ES-RL with learning rates similar to $\alpha_2$. However, the three ES fine-tuning curves in Fig. \ref{fig-two-stage-learning-curves} show that even with a reasonable learning rate, using ES-RL for fine-tuning does not yield desirable results.
Second, directly using PPO for global policy search can be slow and might be trapped in local optima. Given the same computational resources for the first stage learning in the previous example (i.e., 20 node-hours), Fig. \ref{fig-ppo-from-scratch} shows the learning curves of using PPO for global policy search with different node numbers. Two observations are made: a) scaling PPO to have more parallel workers is not beneficial: more parallel workers do not lead to more accurate policy gradient computation and faster convergence; in contrast, the increase communication burden (i.e., $\mathcal{O}(N^2)$) might slow the learning down. b) even for the best performing case (i.e., 1 node), PPO converges to a local minimum and is trapped there for the latter 90\% of the training time (purple shaded area, note x-axis is in log).

\subsection{RLC Control Performance}

\begin{figure}[]
\centering
\includegraphics[width=1.0\linewidth]{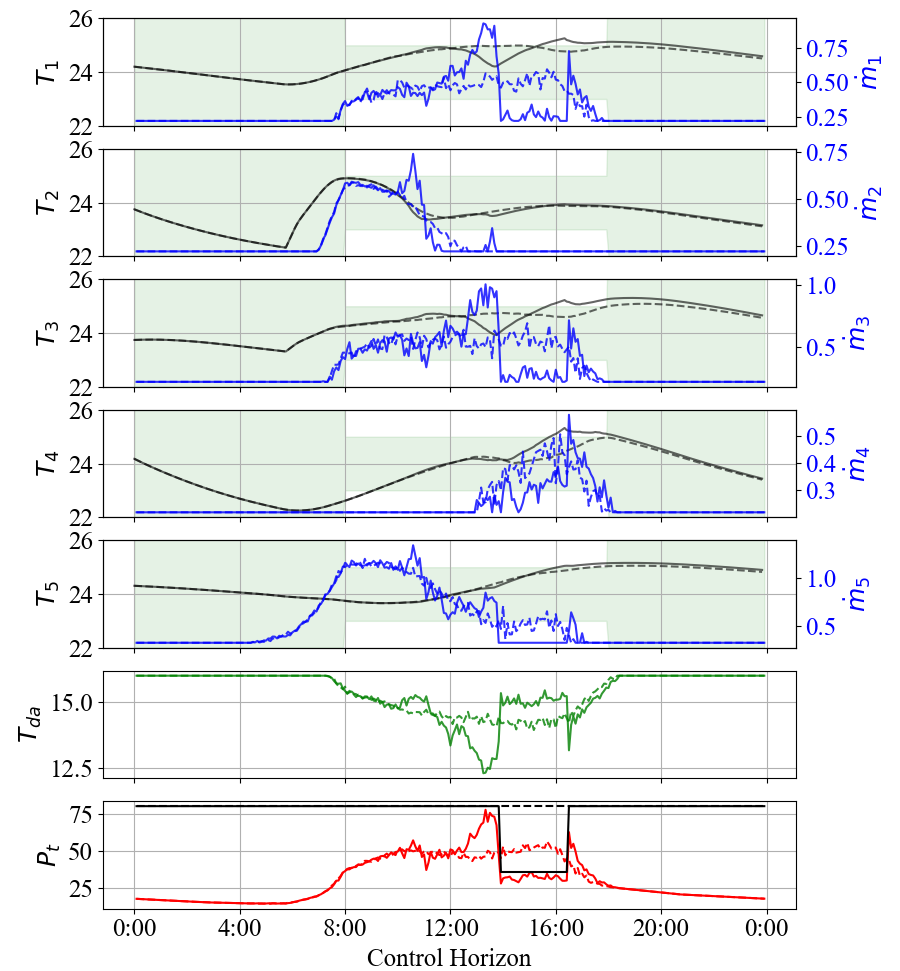}
\caption{Top five plots show the indoor temperature profiles (black) and the air flow rate ($\dot m^i$) profiles (blue) for five zones. Shaded green areas show the temperature comfort band (The band spans [23, 25] and [22, 28] during occupied and unoccupied period, respectively). The sixth plot shows $T_{da}$ and the last plot shows $P_t$ (red) and $\overline{P}_t$ (black). In all plots, dashed lines represent the non-DR day profiles and solid ones show those of the DR day's.}
\label{fig-dr-nodr-comparison}
\end{figure}

To show the effectiveness of the proposed RLC for building control during normal operation and DR events, we simulate two cases where there is/isn't a DR event for a same testing day. Fig. \ref{fig-dr-nodr-comparison} shows the control performance over the 24-hour control horizon for both cases. In the DR case, the building HVAC system is required to limit its power consumption below 36 kW ($\chi=0.3$) between 14:00 and 16:36. The RLC examined is the PPO fine-tuned one (`B' in Fig. \ref{fig-two-stage-learning-curves}).

In Fig. \ref{fig-dr-nodr-comparison}, for the non-DR case (dashed lines), $T_t^i (\forall i, \forall t)$ are kept within the comfort band, and $P_t$ is maintained below $\overline P_t$, with a peak demand of 56.40 kW. On the other hand, in the DR case, upon receiving the DR notification (four hours prior to the event starts), it can be seen that there are pre-cooling activities: first, $T_{da}$ is reduced to accelerate cooling; second, regarding the control of $\dot m_t^i$, we observed a clever learned behavior: the RLC does not mindlessly increase all $\dot m_t^i$ since the fan power will increase drastically due to the cubic term in (\ref{eq-ac-power}); instead, it categorizes zones into three groups and treats them differently: 

1) Zone 4: it faces west and receives great solar heat gain during the DR event. Thus, cooling it \textit{during} a DR event is essential and it is not pre-cooled.

2) Zone 2 and 5: east-facing and the core zone are usually not impacted by the afternoon solar heat gain. The RL agent learned from experience that even without pre-cooling, reducing airflow rates in these zones to minimum during DR events will not jeopardize thermal comfort.

3) Zone 1 and 3: they usually require certain amount of cooling in the afternoon. But they will be the focus of pre-cooling in order to shift the cooling load prior to the event.

During the DR event, $T_{da}$ is increased to reduce chiller power consumption and most of the cooling air flow goes to the west-facing Zone 4, while $T_t^i (\forall i, \forall t)$ are mostly within the comfort band and the load reduction requirement is successfully achieved.

\subsection{Comparison with Model Predictive Controllers}

\begin{figure} []
    \centering
  \subfloat[DR Scenarios\label{1a}]{%
       \includegraphics[width=0.99\linewidth]{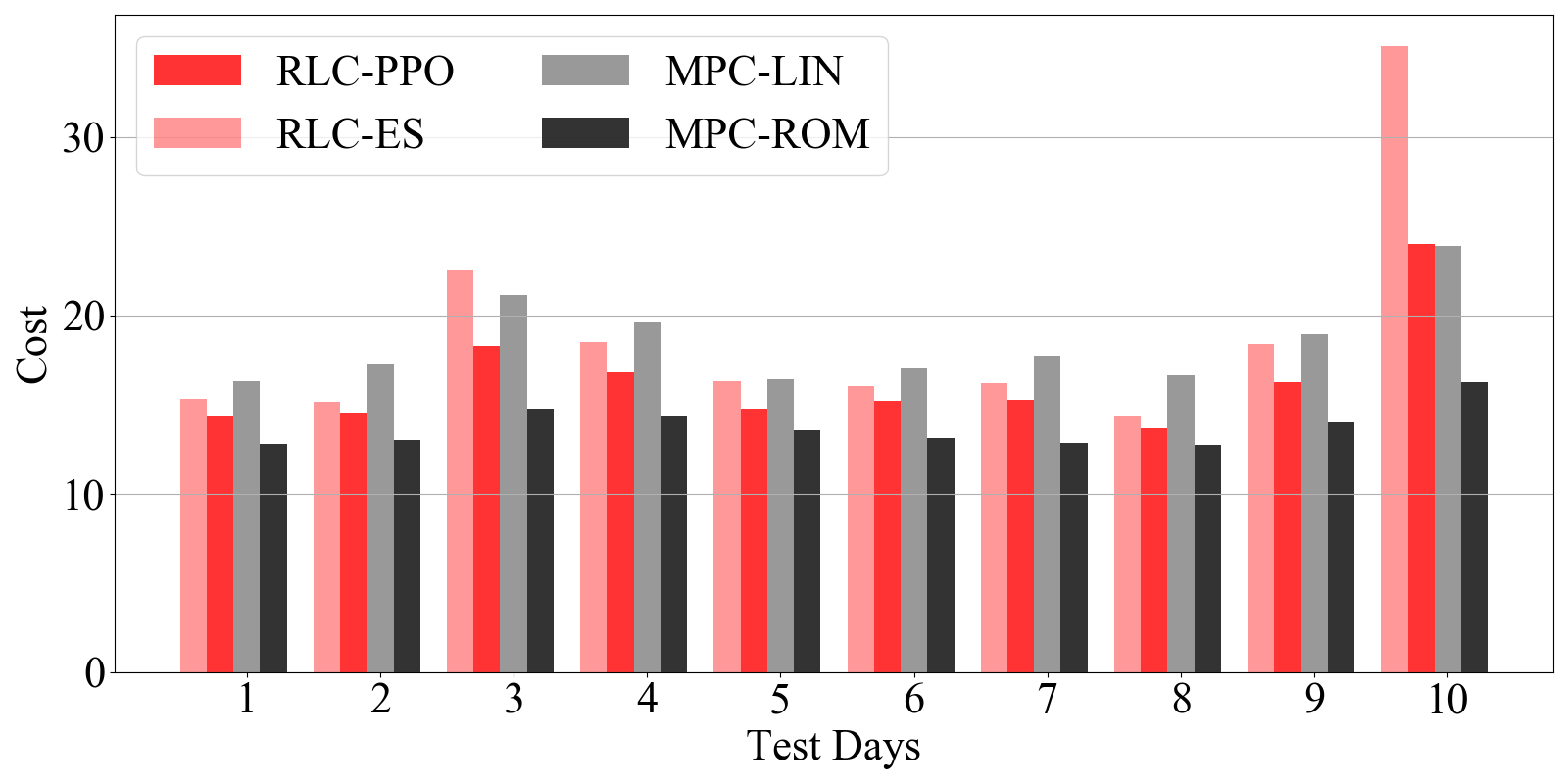}}
    \\
  \subfloat[Non-DR Scenarios\label{1b}]{%
        \includegraphics[width=0.99\linewidth]{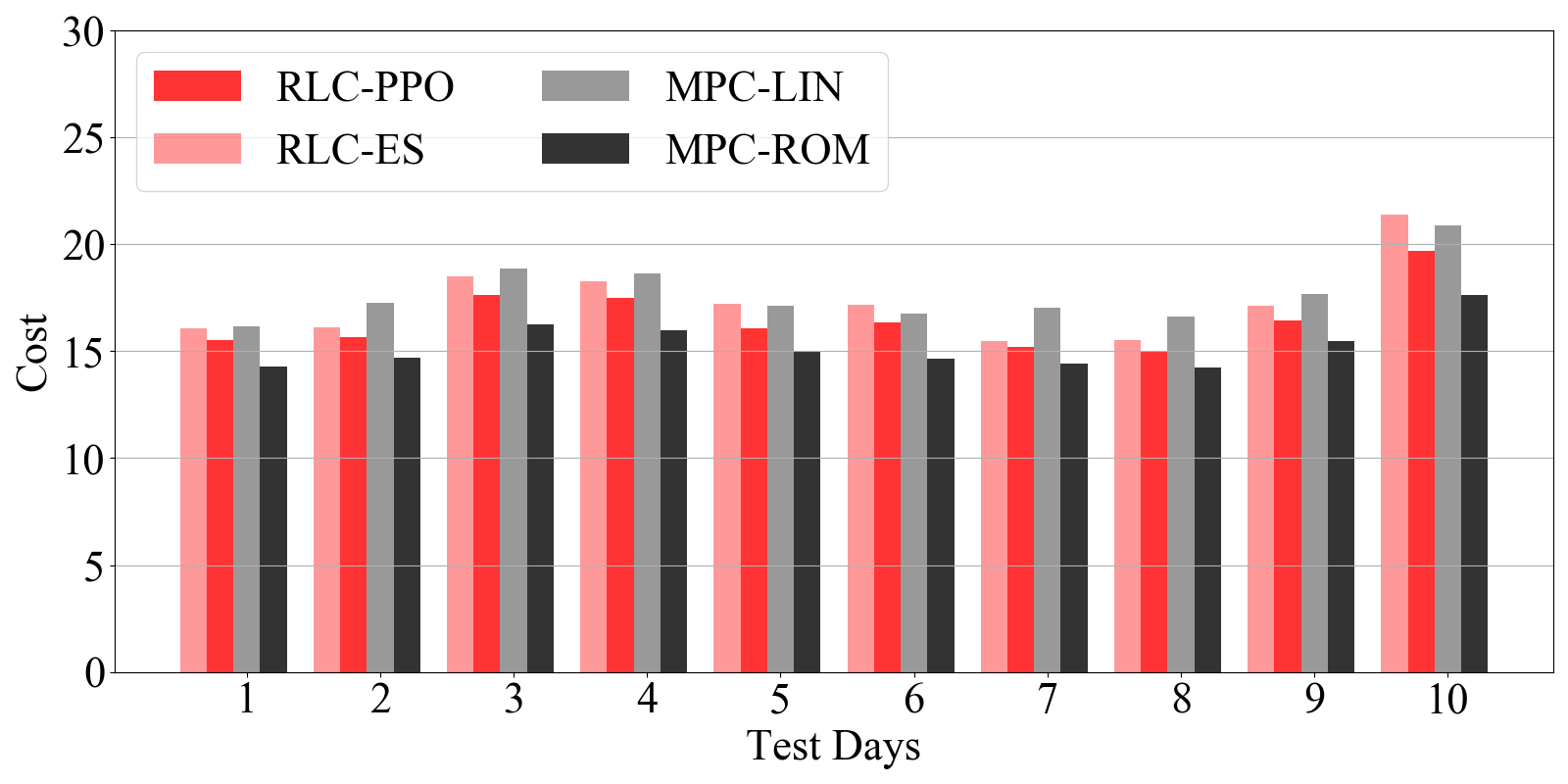}}

  \caption{Comparing RLCs with MPCs.}
  \label{fig-controller-comparison} 
\end{figure}

The ES-RL trained controller (extracted from `A' in Fig. \ref{fig-two-stage-learning-curves}, denoted as `RLC-ES'), PPO fine-tuned RLC (extracted from `B' in Fig. \ref{fig-two-stage-learning-curves}, denoted as `RLC-PPO') and the two MPCs introduced in Section \ref{subsec-mpc} are compared using all ten testing days, and the results are shown in Fig. \ref{fig-controller-comparison} and Table \ref{table-controller-cost-comparison}. Both MPCs are implemented using the $fmincon$ solver in MATLAB with the interior-point algorithm.

The following observations are made:

\begin{enumerate}
    \item Both RLCs can achieve performance similar to or better than that of the MPC-LIN.
    \item The second stage PPO fine-tuning is effective as it reduces the RL-ES's cost by 13.2\% and 4.5\% for DR scenarios and non-DR scenarios respectively.
\end{enumerate}

\begin{table}[]
\caption{Comparison of Average Cost for Test Days}
\centering
\begin{tabular}{c|cccc}
\hline
\textbf{Average Cost}      & \textbf{RL-PPO} & \textbf{RL-ES} & \textbf{MPC-LIN} & \textbf{MPC-ROM} \\ \hline
\textbf{DR}         & 16.31          & 18.79          & 18.50               & 13.75               \\ \hline
\textbf{Non-DR} & 16.50          & 17.28          & 17.70               & 15.26               \\ \hline
\end{tabular}
\label{table-controller-cost-comparison}
\end{table}

It is worth reemphasizing that both MPC-LIN and MPC-ROM use perfect forecasts for exogenous data (i.e., $T_t^{out}$, $Q_t^{solar, i}$ and $Q_t^{int, i}$) and have either approximate or exact prior knowledge on the building being controlled; in contrast, RLC relies only on easily-accessible historical data for control and has zero knowledge on the building model. Overall, this comparison shows that RLC can be a practical and powerful alternative to MPC in real-world building control applications.

\section{Conclusions}

We demonstrated training an RL controller for a grid-interactive multi-zone office building using the proposed global-local policy search scheme.  We showed that the RL controller can optimally control a multi-zone building, under both normal operation and more complicated scenarios such as when providing grid services. Additionally, continuous action space is considered and only easy-to-measure inputs are used for decision making to avoid prohibitive sensor deployment. The experiments reveal that by chaining two different state-of-the-art policy searching paradigms, a control policy can be efficiently trained in the two-stage process. The trained policy, when tested on unseen scenarios, shows proper intelligent control behavior that is entirely learned from the RL agent’s trial-and-error experience collection. Moreover, we showed that the RL controller can achieve a similar performance metrics when compared with two different MPC controllers, whose inputs are perfect forecast information.

In this study, different exogenous data are used in controller training and its performance evaluation (i.e., testing using unseen scenarios), however, we assume the building model is accurate (i.e., perfect system identification). In the future, we will consider the discrepancy between the model used in building simulator and the actual building model, and discuss how to train an effective controller using inaccurate building model.





\bibliographystyle{unsrt}
\bibliography{main}

\end{document}